\documentclass[floats,floatfix,showpacs,amssymb,prl,twocolumn,superscriptaddress,nofootinbib]{revtex4-1}

\newlength{\figw} 
\usepackage{notoccite}

\usepackage{graphicx,epsf, epsfig, amssymb}
\usepackage{bm}
\usepackage{longtable,tabularx}
\usepackage{color}
\usepackage[breaklinks]{hyperref}
\usepackage{amsfonts,amsmath,amssymb,mathrsfs,gensymb}
\usepackage{natbib}
\usepackage{array}
\usepackage{multirow}
\usepackage{rotating,array}
\usepackage{graphicx}
\usepackage{soul}
\usepackage[normalem]{ulem}

\definecolor{linkcolor}{rgb}{0.0,0.3,0.5}
\definecolor{venetianred}{rgb}{0.78, 0.03, 0.08}
\hypersetup{colorlinks=true,linkcolor=linkcolor,citecolor=linkcolor,filecolor=linkcolor,urlcolor=linkcolor}

\newcommand{\sun}{\ensuremath{\odot}}

\def\apj{\rm{ApJ}}
\def\apjl{\rm{ApJ}}

\def\mnras{\rm{MNRAS}}

\def\prd{\rm{Phys.~Rev.~D}}

\def\nat{\rm{Nature}}

\begin{document}

\title{Black hole kicks as new gravitational wave observables}

\author{Davide Gerosa}
\email{d.gerosa@damtp.cam.ac.uk}
\affiliation{Department of Applied Mathematics and Theoretical Physics, Centre for Mathematical Sciences, University of Cambridge, Wilberforce Road, Cambridge CB3 0WA, UK}

\author{Christopher J. Moore}
\email{cjm96@ast.cam.ac.uk}
\affiliation{Institute of Astronomy, University of Cambridge, Madingley Road, Cambridge CB3 0HA, UK}

\pacs{\vspace{-0.15cm}04.25.dg, 04.30.-w, 04.70.Bw, 04.80.Nn}

\date{\today}

\begin{abstract}
Generic black hole binaries radiate gravitational waves anisotropically, imparting a recoil, or kick, velocity to the merger remnant. 
If a component of the kick along the line of sight is present, gravitational waves emitted during the final orbits and merger will be gradually Doppler shifted as the kick builds up.
We develop a simple
prescription to capture this effect in existing waveform models,
showing that future gravitational wave experiments will be able to
perform direct measurements, not only of the black hole kick velocity,
but also of its accumulation profile.  In particular, the eLISA
space mission will measure supermassive black hole kick velocities
as low as $\sim 500$ $\mathrm{km}\,\mathrm{s}^{-1}$, which are expected
to be a common outcome of black hole binary coalescence
following galaxy mergers. Black hole kicks thus constitute
a promising new observable in the growing field of gravitational wave astronomy.
\end{abstract}
\maketitle 

\noindent{\bf \em Introduction.~--~}
Merging black hole (BH) binaries have entered the realm of observational
astronomy. On September 14, 2015, gravitational waves (GWs) emitted
during the inspiral and merger of two stellar-mass BHs of $\sim\!30
M_\sun$ at $z\sim\!0.1$ were detected by the two LIGO detectors
\cite{2016PhRvL.116f1102A}. GW150914 constitutes not only the first
direct detection of GWs, but also the first observation of a stellar-mass BH binary. The identification of supermassive BH
binary candidates has (so far) only been possible through
electromagnetic observations
\cite{2011CQGra..28i4021S,2012AdAst2012E...3D}. The most promising
candidates have been identified as double-core radio galaxies
\cite{2006ApJ...646...49R} and quasars with periodic behaviors
\cite{2008Natur.452..851V,2015Natur.518...74G}. Upcoming GW
observations will revolutionize the field of BH binary astrophysics:
stellar-mass BH binaries will be targeted by a worldwide
network of ground-based interferometers \cite{2015CQGra..32b4001A,2013PhRvD..88d3007A,2010CQGra..27h4007P} while in space the recent success of the LISA pathfinder mission \cite{LPF_ref} has helped paved the way for eLISA \cite{2013arXiv1305.5720C} which will observe hundreds (if not thousands) of supermassive BH binaries out to cosmological redshifts and open the era of multifrequency GW astronomy.

In this {Letter}, we show that the enormous potential of future
GW observations is further enriched by the direct
observability of BH kicks. BH binaries radiate GWs anisotropically which leads to a net emission
of linear momentum and, by conservation of momentum, to a recoil of
the final remnant. This effect has been studied extensively using post-Newtonian and numerical
techniques; see, e.g., Ref.~\cite{2010RvMP...82.3069C} and references therein.
The key findings of these studies are that
the merger of nonspinning BHs can only produce kicks of
$\sim\!170\,\mathrm{km}\,\mathrm{s}^{-1}$ \cite{2007PhRvL..98i1101G},
but that recoil velocities as large as $\sim\!5000$ $\mathrm{km}\,\mathrm{s}^{-1}$
are possible if rapidly rotating BHs with suitable spin orientations
collide
\cite{2007PhRvL..98w1101G,2007PhRvL..98w1102C,2011PhRvL.107w1102L}. These
exceptionally large recoils are commonly referred to as
\emph{superkicks}  and their dynamics can be attributed to
antiparallel spin components in the orbital plane
\cite{1995PhRvD..52..821K}.

\begin{figure*}[t]
\includegraphics[width=\textwidth]{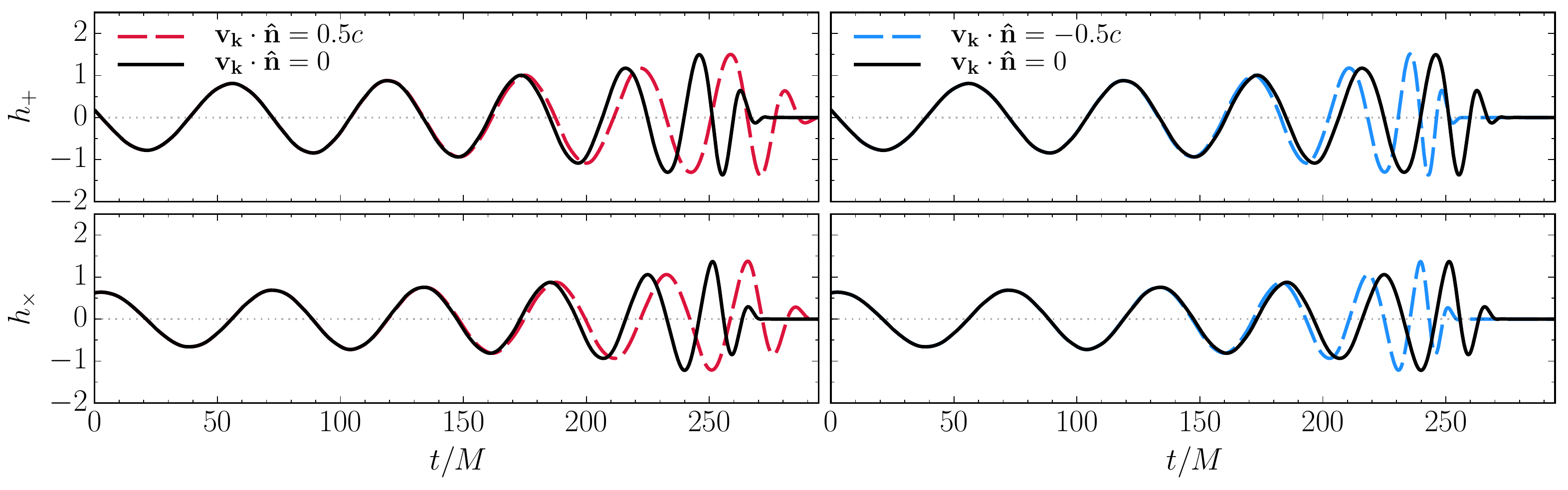}
\caption{GW shift due to BH kicks (artificially exaggerated to
demonstrate the key features). As the
kick velocity builds up during the last few orbits and merger, the
emitted GWs are progressively redshifted (left) or blueshifted
(right), depending on the sign of the projection of the kick velocity
$\mathbf{v_k}$ onto the line of sight $\mathbf{\hat n}$. This is
equivalent to differentially rescaling the binary's total mass in the phase evolution
from $M$ to $M(1+\mathbf{v_k}\cdot\mathbf{\hat n})$. These figures
have been produced by artificially imparting
kicks of $\mathbf{v_k}\cdot\mathbf{\hat n}=\pm 0.5c$ to nonspinning
equal-mass binaries, assuming a Gaussian kick model with $\sigma=60\,M$
[see Eqs.~(\ref{dvdt}) and (\ref{schr}) with $\alpha_{n}=0$ for $n\geq 1$].
}
\label{stretch}
\end{figure*}

BH kicks have striking astrophysical consequences,
especially for supermassive BHs.
Superkicks of $\mathcal{O}(1000)\,\mathrm{km}\,\mathrm{s}^{-1}$ easily exceed the
escape velocity of even the most massive galaxies
\cite{2004ApJ...607L...9M}, and may thus
eject BHs from their hosts \cite{1989ComAp..14..165R}. 
Such ejections would affect the fraction of galaxies hosting central BHs
\cite{2010MNRAS.404.2143V,2015MNRAS.446...38G} and,
consequently, the expected event rates for eLISA \cite{2007MNRAS.382L...6S}.
Even smaller recoil velocities
$\lesssim 500\,\mathrm{km}\,\mathrm{s}^{-1}$ affect the dynamics of galaxy cores by displacing the post-merger BHs for time scales as large as $\sim 10\,\mathrm{Myr}$
\cite{2008ApJ...678..780G,2008ApJ...689L..89K}. BH kicks may lead to a variety of electromagnetic signatures
\cite{2012AdAst2012E..14K}
and observational strategies \cite{2016MNRAS.455..484R,2016MNRAS.456..961B}
have recently been proposed for their detection. Candidates are present, but their nature is debated (see Refs.~\cite{2012AdAst2012E..14K,2014MNRAS.445..515K} and references therein) and, overall, BH kicks remain elusive.

If GW observations of a BH binary provide accurate measurements of the component masses and spins,
it is, in principle possible to use numerical relativity
(NR) results to infer
the kick that the binary should have received around merger (this was  not possibile for GW150914 \cite{2016arXiv160203840T}).
Such an approach, however, would be of \emph{indirect} nature and
crucially relies on the validity of the assumptions in
the numerical modeling process. For instance, it would {\em not}
provide an additional consistency check of the predictions of
general relativity (GR).
As argued here, it is possible instead to \emph{directly} measure
BH kicks from the GW signal alone. If the kick is directed towards
(away from) Earth, then the latter part of the waveform will be blueshifted
(redshifted) relative to the early part. Roughly speaking, different,
Doppler-shifted mass parameters
would be inferred from the inspiral and ringdown
parts of the signal if analyzed separately. More precisely, by
observing the differential Doppler shift throughout the signal, one
can directly measure the change in speed of the system's center of
mass as a function of time. 

\medskip

\noindent{\bf \em Doppler mass shift.~--~}
In the absence of a mass or length scale in vacuum
GR, the GW frequency $f$ enters the binary dynamics exclusively in the
dimensionless form $f M$, where $M$ is the total mass of the binary (hereafter
$G=c=1$).
This scale invariance implies a complete degeneracy between a
frequency shift and a rescaling of the total mass of the system.
For example, the cosmological redshift $z$ of a BH binary merely enters
in the predicted GW emission through a rescaling of the total mass
by a factor $(1+z)$ and, hence, GW observation of the binary
only measures the  combination $M(1+z)$\cite{1987GReGr..19.1163K}. BH kicks produce a similar effect: at linear order, the motion of the center
of mass shifts the emitted GW frequency by a factor
$1+\mathbf{v_k}\cdot\mathbf{\hat n}$ while leaving the amplitude unaffected ($\mathbf{v_k}$ is the
kick velocity with magnitude $v_k$ and the unit vector $\mathbf{\hat n}$ denotes the direction of the line of sight from observer to source). There is, however, one crucial difference: while
cosmological redshift homogeneously affects the entire signal,
a frequency shift due to BH kicks gradually accumulates during the
last orbits and merger. This point is illustrated in Fig.~\ref{stretch}: 
as a kick is imparted to the merging BHs, the emitted GWs are
progressively blue- or redshifted. 
The frequency of the signal changes as if the mass
of the system was varied from $M$ in the early inspiral to $M(1+\mathbf{v_k}\cdot\mathbf{\hat n})$ 
by the end of the ringdown.

The detectability of this effect can be estimated using the
following back-of-the-envelope argument. Imagine breaking a BH binary waveform
into two parts: inspiral and ringdown,  $h(t)=h_{i}(t)+h_{r}(t)$.
For simplicity, assume that the kick is imparted instantaneously
at merger so that only $h_r$ is affected. 
Let $M_i$ and $M_r$, respectively, denote the total binary mass as measured from $h_i$ and $h_r$ alone.
Neglecting the energy radiated in GWs --this effect is not
negligible in magnitude, resulting in a reduction of the mass by $\sim 5~\%$,
but can be estimated accurately from the waveform and thus
be accounted for--, the
effect of a kick is to Doppler shift the final mass according to $M_r=M_i
(1+\mathbf{v_k}\cdot\mathbf{\hat n})$.
The inspiral part $h_i$ of the GW signal generally contains a larger fraction of the signal-to-noise ratio (SNR) than the ringdown
part $h_r$, so the detectability of the kick will be
limited by the measurement of $M_r$: kicks of magnitude $v_k$ can
be detected if $M_r$ is measured with a fractional accuracy of
$\lesssim v_k /c$ ($\sim 1\%$ for a superkick along the line
of sight). 
The ringdown waveform can be modeled using the least damped quasinormal mode for a Schwarzschild BH \cite{2006PhRvD..73f4030B}
$h_{r}(t) \simeq A \exp(-0.089 t/M_{r})\sin(0.37 t/M_{r})$
which gives a squared SNR
\begin{equation}\label{eq:rho} \rho_{r}^{2}=\frac{1}{S_{n}}\int_0^\infty h_r(t)^{2} \,\mathrm{d}t \simeq\frac{2.66 M_r A^{2}}{S_{n}}  \,,\end{equation}
assuming white noise in a detector with power spectral density (PSD) $S_{n}(f)=S_n={\rm const}$. The error on the measurement of $M_r$ can be estimated using the linear signal approximation, 
\begin{equation}
\label{eq:deltaM}
 \left(\frac{1}{\Delta M_r}\right)^{2}=\frac{1}{S_{n}}\int_0^\infty\left(\frac{\partial }{\partial M} h_{r}(t)\right)^{2} \,\mathrm{d}t\simeq\frac{25.6 A^2}{ M_r S_{n}}\,, 
 \end{equation}
Therefore, the fractional error on $M_r$ is given by
\begin{equation}
\label{eq:estimate}
\frac{\Delta M_r}{M_r} \simeq \frac{0.322}{\rho_{r}}
\,.
\end{equation}
This back-of-the-envelope argument suggests that kicks along the
line of sight with magnitude $v_k\sim0.003 c \simeq
900\,\mathrm{km}\,\mathrm{s}^{-1}$  can be measured with GW
observations if the SNR in the ringdown is $\rho_r\sim 100$. 
Direct detection of BH kicks will be very challenging, if not impossible,
with current ground-based detectors. For instance, the
rather loud event GW150914 has a ringdown SNR
$\rho_r\sim 5$ \cite{2016PhRvL.116v1101A}, which would only allow
us to measure unrealistically large
kicks $v_{k}\sim0.06c$. On the other
hand, BH kicks are very promising observables for space-based
detectors, where SNRs in the ringdown can reach
$\rho_r\sim 10^3$ \citep{2013LRR....16....7G}. This will allow for
measurements of supermassive BH kicks with magnitude as low as $v_k\sim
100\,\mathrm{km}\,\mathrm{s}^{-1}$, which are expected to be ubiquitous
\cite{2012PhRvD..85l4049B,2012PhRvD..85h4015L}.
The detectability of the kick is governed by the ringdown part of the SNR $\rho_r$, which has also been found to be important  to detect the GW memory effect (see Ref.~\cite{2009JPhCS.154a2043F} where kicks are also mentioned) and test the Kerr hypothesis via BH spectroscopy \cite{2006PhRvD..73f4030B}. 
\medskip

\noindent{\bf \em Kicked waveforms.~--~}
In order to investigate the detectability of BH kicks more
quantitatively, we need a waveform model that captures the
cumulative frequency shift they introduce.
Doppler shifts due to BH kicks can be straightforwardly incorporated into any
preexisting waveform model (which does not include the kick) by substituting $M\to M\times[1+v(t)]$ in the phase evolution,
where $v(t)$ is the projection of the center-of-mass velocity due to the kick
onto the line of sight. 
Here, we only consider the nonrelativistic
Doppler shift; relativistic corrections 
enter at the
order $\mathcal{O}(v_k)^2 \lesssim 10^{-4}$, well below the
magnitude relevant for our analysis.
The profile $v(t)$ is taken such that $v(t)\rightarrow 0$ as
$t\rightarrow -\infty$ and $v(t)\rightarrow\mathbf{v_k}\cdot\mathbf{\hat
n}$ as $t\rightarrow \infty$. A common observation in NR simulations
is that the kick is imparted over a time $2\sigma \sim 20 M$ centered
on the merger, at a rate $\mathrm{d}v/\mathrm{d}t$ which is approximately of Gaussian shape
\cite{2008PhRvD..77l4047B,2008PhRvD..77d4028L},
possibly with some deceleration after merger (\!\emph{antikick}) \cite{2007PhRvL..99d1102K,2010PhRvL.104v1101R}. In contrast to the kick speed, relatively little is known regarding the kick profile beyond these qualitative observations. We therefore adopt a flexible model for the kick profile. We expand $\mathrm{d}v/\mathrm{d}t$ according to
\begin{align}
&\frac{\mathrm{d}}{\mathrm{d}t}v(t)=\mathbf{v_k} \cdot \mathbf{n}\frac{ \sum_n  \alpha_n  \phi_n(t)}
{\int_{-\infty}^{\infty} \sum_{n} \alpha_n  \phi_n(t) \,\mathrm{d}t} \,,\label{dvdt}
\\
&\phi_n(t)\!=\!\frac{1}{\sigma\sqrt{2^n n!\sqrt{\pi}}} \exp\!\left(-\frac{(t-t_{c})^{2}}{2\sigma^{2}}\right)H_{n}\!\left(\frac{t-t_{c}}{\sigma}\right), \label{schr}
\end{align}
where $H_n$ are the Hermite polynomials, $t_c$
is the time of coalescence, $\sigma$ controls the duration over
which the kick is accumulated and the $\alpha_n$ weigh the various
components. The functions $\phi_n(t)$ constitute a complete basis (they are actually the familiar solutions for the quantum harmonic oscillator) and so they can model all possible kick profiles. This basis is particularly appealing, because
the first two terms $n=0,1$ model Gaussian acceleration profiles and {antikicks}, respectively.  The case $\sigma=0$
and $\alpha_n=0$ for $n\geq1$ corresponds to a kick instantaneously
imparted at $t_c$, as assumed in the back-of-the-envelope argument
presented above.  We have tested this prescription against 200 NR
waveforms from the public Simulating eXtreme Spacetimes catalog \cite{2013PhRvL.111x1104M},
finding that the radiated-momentum profiles obtained from integrating
the $l\leq 6$ modes of the  Newman-Penrose scalar $\Psi_4$ are well
approximated by the first two terms of the expansion of
Eqs.~(\ref{dvdt}) and (\ref{schr}). For systems with  kicks above $500\,\mathrm{km}\,\mathrm{s}^{-1}$, residuals  in  $v_k$   
are less than 17\% in all cases, and typically less than 4\% \cite{followup}.

\begin{figure}
\includegraphics[width=\columnwidth]{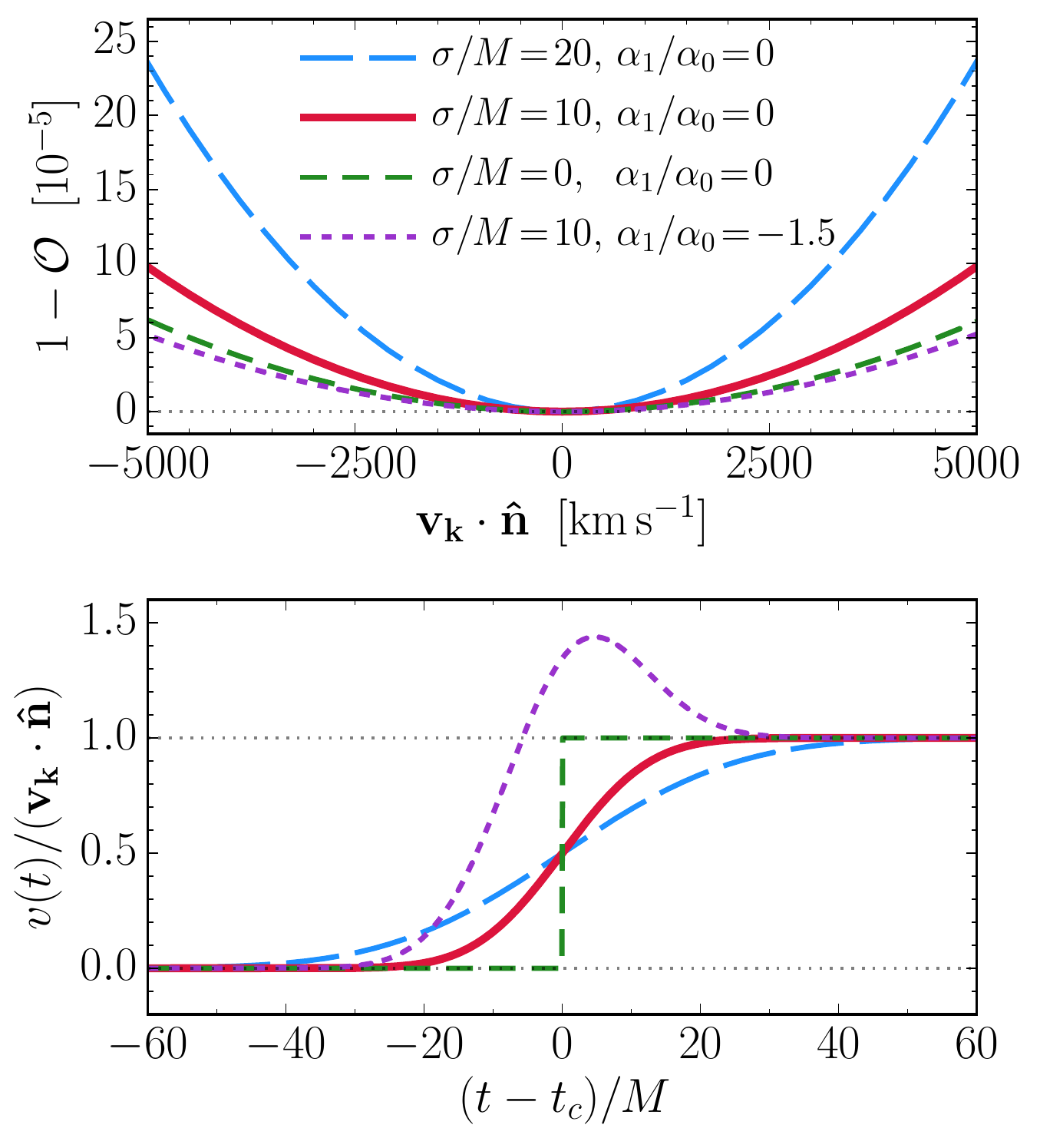}
\caption{Mismatches introduced by BH recoils. The top panel shows
the mismatch $1-\mathcal{O}$ between (i) a standard waveform of
equal-mass nonspinning BH binaries of total mass $M$ and (ii) a
`kicked' waveform which includes the Doppler-shifting effects of a velocity profile $v(t)$.
Each line corresponds to a different kick profile
$v(t)$, as shown in the bottom panel. All models shown here assume
$\alpha_n=0$ for $n\geq 2$. The $\sigma=10M$, $\alpha_1/\alpha_0=0$
model (solid line) is used in Fig.~\ref{SNRvsmatch}.
}
\label{shape}
\end{figure}

\begin{figure*}[!t]
\includegraphics[width=\textwidth]{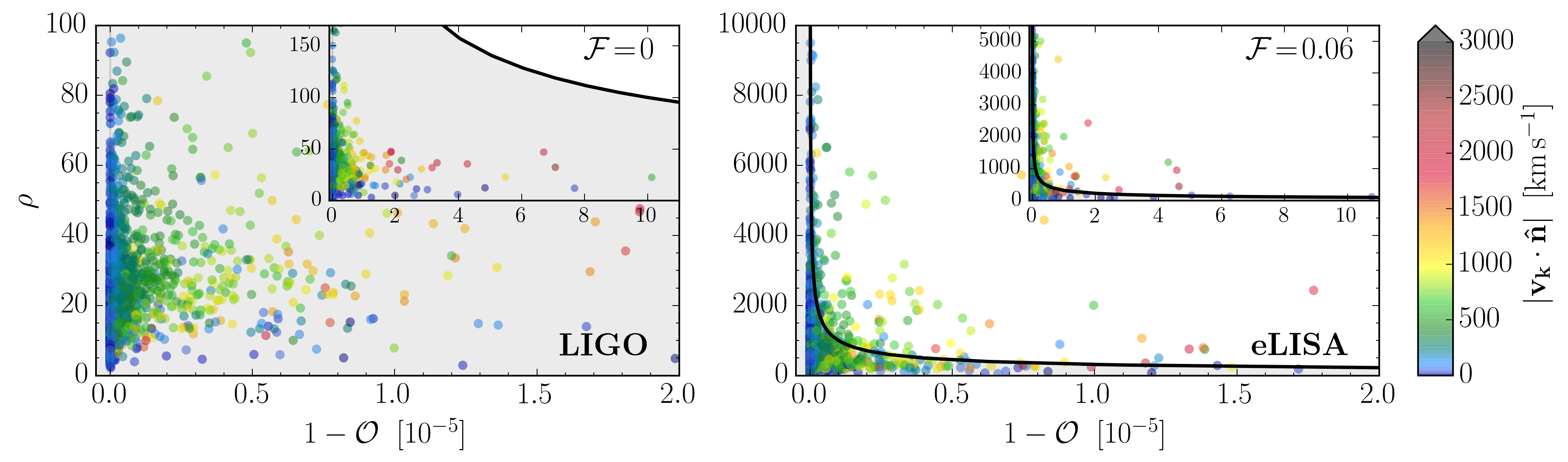}
\caption{Detectability of BH kicks with  LIGO (left) and eLISA
(right). For each simulated source we compute the overlap $\mathcal{O}$
between standard and `kicked' waveforms, and compare it with the
SNR $\rho$. Kick velocities --here encoded in the color bar-- are
imparted using NR fitting formulas. BH kicks are detectable for the
fraction $\mathcal{F}$ of the sources above the black line, $\mathcal{O}<
1-\rho^{-2}$. eLISA results have here been generated with the ``N2A5L6'' PSD of Ref.~\cite{2016PhRvD..93b4003K}. }
\label{SNRvsmatch}
\end{figure*}

For a given waveform approximant, GW detector, and binary parameters,
we generate two signals: a standard waveform $h_0(t)$ 
and a second ``kicked'' waveform $h_k(t)$.
The two waveforms can be compared by calculating
their overlap
 \begin{align}
\mathcal{O} = \max_{t_c,\phi_c} \frac{(h_0|h_k)}{\sqrt{(h_0|h_0)(h_k|h_k)}}\,,
\end{align}
where $(h_0|h_k)$
 is the noise-weighted inner product \cite{2015CQGra..32a5014M}
 and $t_c$ ($\phi_c$) is the time (phase) of coalescence. Approximately, two waveforms are distinguishable (and the kick detectable) if $\mathcal{O}\lesssim 1-\rho^{-2}$ \cite{2008PhRvD..78l4020L}, where $\rho=\sqrt{(h_0| h_0)}$ is the SNR (of the full waveform). This assumes the kick is not degenerate with other parameters, which is expected as the kick mostly affects the ringdown and not the entire signal. 

This procedure is illustrated in Fig.~\ref{shape} using a
simple controlled experiment. We consider 6 inspiral cycles, merger
and ringdown of an equal-mass nonspinning BH binary (a similar
setup to that used in Fig.~\ref{stretch}). For simplicity, and to ensure
that the results are not detector specific, the overlaps have been
computed using a flat PSD. Artificially imposed recoils of $\sim
1000$ $\mathrm{km}\,\mathrm{s}^{-1}$ introduce mismatches
$(1-\mathcal{O})\sim 10^{-5}$.
Kicks are more likely to be detected
if they are imparted
over a longer period of time (i.e. larger $\sigma$) because dephasing
starts to occur earlier in the inspiral (this effect can be seen
in Fig.~\ref{stretch} where a larger value of $\sigma = 60\,M$ was
used). Note that the overlaps are approximately symmetric with respect to the transformation
$\mathbf{v_k}\to -\mathbf{v_k}$, i.e. blueshifts and redshifts are
equally detectable. This property can be shown to hold exactly
at linear order in $v_k$ \cite{followup}.

We next explore more realistic scenarios by using NR fitting formulas
to predict the kick velocity. For this purpose, we generate two
BH binary populations for the LIGO and eLISA detectors.  LIGO
(eLISA) sources were selected randomly from the
following distributions: uniform total mass 
$M \in [10M_\odot,100 M_\odot]$ ($[10^5M_\odot,10^6 M_\odot]$) and mass
ratio $q\in[0.05,1]$; uniform  dimensionless spin magnitudes
$\chi_1,~\chi_2\in[0,1]$;  isotropic inclination and spin directions  at a reference GW frequency
$f_{\rm ref}=20$ Hz ($2$ mHz); isotropic sky location; sources are distributed homogeneously in
comoving volume with comoving distance $D_c \in [0.1\,{\rm Gpc}, 1  \,{\rm Gpc}]$ ($[1\,{\rm Gpc}, 10\,{\rm Gpc}]$)  assuming the Planck cosmology \cite{2015arXiv150201589P}. 
We use the LIGO \mbox{``Zero-Det-High-P''} PSD of Ref.~\cite{LIGOcurve} with lower cutoffs $f_{\rm low}=10$ Hz, and the two possible 
eLISA PSDs ``N2A5L6'' and ``N2A1L4'' of Ref.~\cite{2016PhRvD..93b4003K} with $f_{\rm low}=0.3$ mHz (the former being more optimistic; for simplicity, we neglect the spacecraft orbital motion which can be separately accounted for).
For each binary, we estimate the kick velocity using the fitting
formula summarized in \cite{2016arXiv160501067G}.
In order to return accurate estimates,
the kick formula requires as input the BH spin parameters
at separations $r\sim10M$, comparable
to the initial separations of the NR simulations used in the formula's
calibration.
Otherwise, resonant effects \cite{2004PhRvD..70l4020S}
are not adequately accounted for and lead to erroneous kick magnitudes \cite{2010ApJ...715.1006K}.
We  bridge the separation range between $f_{\rm ref}$ and $r = 10~M$ using the orbit-averaged post-Newtonian evolution code of \cite{2016arXiv160501067G}.
The NR fitting formula then provides expressions for the kick components parallel
and orthogonal to the binary orbital angular momentum $\mathbf{L}$:
$v_\parallel$ and $v_\perp$. The projection of the kick velocity
along the line of sight is given by
\begin{align}
\mathbf{v_k}\cdot \mathbf{\hat n} = v_\parallel \cos\Theta \cos\iota - v_\perp \cos\Theta' \sin\iota\,,
\end{align}
where $\cos\iota=\mathbf{\hat L}\cdot \mathbf{\hat n}$ is the cosine
of the inclination at $r=10M$, $\Theta$ is related to the direction
of the orbital-plane components of the spins at merger
\citep{2008PhRvD..77l4047B,2009PhRvD..79f4018L}, and $\Theta'$ sets
the direction of the orbital-plane component of the kick \cite{followup}.
In practice, both $\Theta$ and $\Theta'$ depend on the initial
separation of the binary in the NR simulations. While the $\Theta$
dependence has been studied extensively in the literature
\citep{2008PhRvD..77l4047B,2009PhRvD..79f4018L}, the impact of
$\Theta'$ and its relation with $\Theta$ have, to our knowledge,
not yet been explored.
In the following, both angles are drawn uniformly in $[0,\pi]$. For
each system, we generate two waveforms, $h_0$ and $h_k$, using the
inspiral-merger-ringdown approximant \mbox{``IMRPhenomPv2''} of Refs.~\cite{2014PhRvL.113o1101H,2016PhRvD..93d4007K,2016PhRvD..93d4006H}
which accounts for spin precession. We have verified our results for the
overlaps are insensitive to the choice of the waveform approximant,
even when non-precessing models are used.
In the following, we assume a ``Gaussian'' kick model, described
by $\alpha_n=0$ for $n\geq1$ and $\sigma=10M$ (solid curve in
Fig.~\ref{shape}); cf. Ref.~\cite{2008PhRvD..77l4047B}.

Our results are summarized in Fig.~\ref{SNRvsmatch}. As suggested
by our previous argument, none of the LIGO sources have mismatches
high enough to detect the kick. The eLISA case is 
different: $\sim 1\%$-$6\%$ (depending on the PSD) of the simulated sources have $\mathcal{O}<
1-\rho^{-2}$ and therefore present detectable BH kicks.  Kicks with
a projected magnitude $\mathbf{v_k}\cdot \mathbf{\hat n}\gtrsim 500$
$\mathrm{km}\,\mathrm{s}^{-1}$ at $\rho\gtrsim 1000$ will be
generically observable, but even some of the lower kicks with
$\mathbf{v_k}\cdot
\mathbf{\hat n} \sim 100$ $\mathrm{km}\,\mathrm{s}^{-1}$ may
be accessible. In the fortunate case  of a \emph{superkick} directed
along the line of sight ($|\mathbf{v_k}\cdot \mathbf{\hat n}|\sim
3000$ $\mathrm{km}\,\mathrm{s}^{-1}$), the effect may be so prominent
to be distinguishable at SNRs as low as $\rho\sim 50$. 
As eLISA is expected to measure up to $\mathcal{O}(100)$ BH binaries per year \cite{2013arXiv1305.5720C,2016PhRvD..93b4003K}, our study suggests that $\sim 6~{\rm yr}^{-1}$ ($\sim 30$ in total for a 5-yr mission lifetime) sources may present detectable kicks.
Although
more realistic astrophysical modeling is needed to better quantify
this fraction, our simple study shows that direct detection of BH
recoils is well within the reach of eLISA. Third-generation
ground-based detectors will also present promising opportunities:
repeating the calculations of the LIGO population of binaries but
observed with ET (assuming the ``ET-D-sum'' PSD of Ref.~\cite{ETcurve}, with $f_{\rm{low}}=1$ Hz)
we find $\sim 5\%$ of binaries possess detectable kicks.

GW observations not only have the potential to measure the magnitude
of the BH kick, but also the details of how the velocity accumulates
with time. By expanding $v(t)$ according to Eqs.~(\ref{dvdt}) and (\ref{schr}),
one can take the kick model parameters  $\mathbf{v_k}\cdot \mathbf{\hat
n}$, $\sigma$ and $\alpha_n$ to be free parameters of the waveform
model, and treat them on an equal footing with  masses, spins,
inclination angles, etc. Consider, for example, a \emph{golden system} at $\rho=10^4$
with component masses of $1.3\times 10^{6}\,M_{\sun}$ (chosen to
maximize the mismatch caused by the kick),
misaligned extremal spins and inclination such that $\mathbf{v_k} \cdot \mathbf{\hat
n}\sim 5000\,\mathrm{km}\,\mathrm{s}^{-1}$ km. A Fisher matrix
calculation of the intrinsic parameters of this binary suggests
that eLISA will be capable of measuring the kick velocity with precision
$\Delta v_k \sim 200\,\mathrm{km}\,\mathrm{s}^{-1}$, the kick
duration with precision $\Delta \sigma \sim 1\,M$ and the presence
of an {antikick} at the level of $\Delta (\alpha_{1}/\alpha_{0})\sim
0.1$ (considering a two-component kick model, i.e. $\alpha_n=0$
for $n\geq 2$) \cite{followup}.
This Fisher matrix analysis revealed no strong degeneracies between the kick and other parameters, thus further justifying our previous use of the overlap as a detectability criterion for the kick.
Finally, note that \emph{superkicks} have $v_\parallel \gg v_\perp$
so that face-on or face-off binaries $|{\mathbf{\hat L}\cdot \mathbf{\hat n}}|\sim 1$
generate the largest velocity components along the line of sight
and, hence, are most favorable for a direct kick measurement. 
\medskip

\noindent{\bf \em Conclusions.~--~}
 BH kicks leave a clear imprint on the GW waveform emitted during
 the late stages of the inspiral, merger and ringdown
 of BH binaries. eLISA and, likely, third-generation ground-based detectors
 will be able to \emph{directly} detect the presence of a
 kick from the distortion of the waveform for a significant fraction
 of the binaries observed. By comparing the directly measured kicks
 (both magnitude and profile) to the NR kick predictions for a binary
 with measured masses and spins, it will be possible to verify whether
 linear momentum is radiated as predicted by GR. Much like the
 Hulse-Taylor pulsar provided the first evidence that GWs carry away
 energy in accordance with the expectation of GR, and GW150914
 provided the first direct evidence of the GWs themselves
 \cite{2016PhRvL.116f1102A}, a direct measurement of a BH kick
 will provide the first direct evidence for the linear momentum carried
 by GWs.
\medskip

\noindent{\bf \em Acknowledgments.~--~} We thank Ulrich Sperhake, Jonathan Gair, Enrico Barausse, Emanuele Berti, Christopher Berry, Vitor Cardoso, Alberto Sesana, Nathan Johnson-McDaniel and the anonymous referees for fruitful discussions and suggestions. This document has been assigned LIGO document reference LIGO-P1600118. D.G. is supported by the UK STFC and the Isaac Newton Studentship of the University of Cambridge. C.M. is supported by the UK STFC. Partial support is also acknowledged from the H2020 ERC Consolidator Grant ``Matter and strong-field gravity: New frontiers in Einstein's theory'' Grant Agreement No. MaGRaTh--646597, the H2020-MSCA-RISE-2015 Grant No. StronGrHEP-690904, the STFC Consolidator Grant No. ST/L000636/1, the SDSC Comet and TACC Stampede clusters through NSF-XSEDE Award No.~PHY-090003, the Cambridge High Performance Computing Service Supercomputer Darwin using Strategic Research Infrastructure Funding from the HEFCE and the STFC, and DiRAC's Cosmos Shared Memory system through BIS Grant No.~ST/J005673/1 and STFC Grants No.~ST/H008586/1 and ST/K00333X/1. 
\bibliography{directkick}
\end{document}